\documentstyle[epsfig,aps,prl,preprint]{revtex}
\tightenlines
\begin{document}

\title{Relativistic nuclear structure effects in $(e,e'\vec{p})$}

\author{J.M. Ud\'{\i}as and Javier R. Vignote}
 
\address{Departamento de F\'{\i}sica At\'omica, Molecular y Nuclear\\
Facultad de Ciencias F\'{\i}sicas,
Universidad Complutense de Madrid, 
 E-28040 Madrid, Spain}

\maketitle

\begin{abstract}
 Results for recoil nucleon induced polarization
 for $(e,e'\vec{p})$
 are presented using various approximations for 
the relativistic nucleonic current, at the kinematics of a
recent experiment at Bates. We see that the 
dynamical relativistic effects improve the agreement
with the  data. We make predictions for the induced
normal polarization and responses for
  TJNAF  89-033  and Mainz A1/2-93 experiments.
\end{abstract}

\vspace*{1cm}

\section{Introduction}
Several  experiments have been proposed or have been carried
out to measure the polarization of the ejected nucleon in
$(e,e'\vec{p})$ reactions \cite{Woo,Glasshauser,MAMI}. 
In this way, new sets of
polarization response functions can be isolated
\cite{Boffi,donRask,Giusti,vanorden}. 

If we write the cross section for the coincidence $(e,e'\vec{p})$ 
reaction in terms of recoil
nucleon polarization dependent and independent 
terms, we have \cite{Boffi,Giusti,vanorden}:
\begin{equation}
\frac{d^3\sigma_s}{d\epsilon_e d\Omega_e d\Omega'}=\frac{\sigma_0}{2}\left[1+
\vec{P}\cdot\vec{\sigma}\right],
\label{cross}
\end{equation}
where $\epsilon_e$ is the scattered electron energy, $\sigma_0$ is the
unpolarized cross section, $s$ denotes the nucleon spin
projection upon $\vec{\sigma}$, and $\vec{P}$ is the induced
polarization. Each of these observables can be written in terms
of response functions
 that
are bilinear combinations of the nuclear electromagnetic current
operator
 \cite{Boffi,donRask,vanorden}.
If  the electron beam is unpolarized and the experiment
is performed in coplanar kinematics ($\phi'=0,\pi$), the relationship between
the nuclear responses and the cross section is given by:
\begin{eqnarray}
\label{responses}
\frac{d^3\sigma_s}{d\epsilon_e d\Omega_e d\Omega'} & = &
\frac{E'|\vec{P}'|}{2 (2\pi^3)} 
\left[\frac{d\sigma}{d\Omega_e}\right]_{Mott}  \times   \\ \nonumber
& & \left\{
 V_L (R_L+R^n_L \hat{S}_n)+
 V_T (R_T+R^n_T \hat{S}_n)+ \right.  \\ \nonumber
& & \left. \cos\phi' V_{TL} (R_{TL}+R^n_{TL} \hat{S}_n)+
\cos 2\phi' V_{TT} (R_{TT}+R^n_{TT} \hat{S}_n)\right\} \nonumber
\end{eqnarray}
The kinematical factors are $V_L=\lambda^2$,
$V_T=\lambda/2+\tan^2\theta_e/2$, $V_{TT}=\lambda/2$, 
$V_{TL}=\lambda\sqrt{\lambda+\tan^2\theta_e/2}$ and
$\lambda=1-(\omega/|\vec{q}|)^2$ where $\omega$ and $\vec{q}$
are the energy and momentum transfer in the reaction, $\theta_e$
is the electron scattering angle and $E'$, $|\vec{P}'|$ are
the energy and momentum of the ejected nucleon.
Hence, for coplanar kinematics, {\em i.e.}, when the
ejected nucleon lies within the
electron scattering plane $\vec{P}$, 
the net ejectile polarization
for an unpolarized beam or induced
polarization ($P_n$), is normal
to the scattering plane.
In the one-photon exchange approximation $P_n$ 
is zero  when no final state interactions (FSI)  between the ejected
nucleon and the residual system are
considered \cite{Boffi,donRask,vanorden}.
Thus, $P_n$  is an observable well suited to  study 
FSI effects in nuclear systems and measurements of $P_n$ at 
different $Q^2$ would give information about the onset 
of nuclear transparency.  If nuclear transparency
is present at certain $Q^2$ value, that is, if FSI 
effects are quenched, we would see a  decrease of
 $P_n$ what would be a  clear
signature of nuclear transparency  free from the
ambiguities on the 
occupancies of the
shells under study \cite{trasp}.

 The first analysis of the experiment
performed at BATES by Woo {\em et al.} that measured
$P_n$ in $^{12}C(e,e'\vec{p})$
\cite{Woo}  was made in a non relativistic framework.
Other non relativistic results for this experiment 
were recently presented  in reference \cite{Ryckebush}.
The non relativistic approach to $(e,e'\vec{p})$ is based on
the impulse approximation, {\em i.e.}, assuming the
one-photon exchange picture in which the single
photon interacts only with the nucleon that is detected\cite{Boffi}.
If no FSI are considered, the
ejected nucleon is described by a plane wave (plane
wave impulse approximation or PWIA). FSI are taken into
account using potentials that distort the final
nucleon wave function (distorted wave impulse
approximation or DWIA) \cite{Boffi}.
The non relativistic analyses found a 
 a systematic underestimation of $P_n$ of around
10\% at the best\cite{Woo}.

There are two main sources of an induced nonzero normal polarization
in proton knockout reactions caused by the interaction with the 
residual nucleus in the final state.
 One is due to the absorption, that is,
the flux lost into inelastic channels,  
  parametrized in the 
imaginary part
of the optical potential. 
 Semi-classically,
for scattering on a given side of $\vec{q}$,
  ejecting  a nucleon
from  the front or the rear face of the nucleus would select
out different directions of
the angular momentum $\vec{l}=\vec{r}\times\vec{p}$
and, as the
absorption depends on how much  the  nucleon
travels in the nuclear medium 
before being detected, the effect is a net induced polarization
due to  absorption.
This is well known from hadronic
reactions and is named as the Maris Effect or Newns Polarization \cite{Jacob}.
It is, however, a small source of $P_n$\cite{Woo}. 
The bulk of the induced polarization is primarily due to the real part
of the spin-orbit potential, that parametrizes the explicitly spin-dependent
terms in the optical potential\cite{Woo}. 

Since  the spin is a property
 intrinsically related to  relativity,
one may {\em a priori} consider that a relativistic approach
 is better suited to describe nucleon polarization
 observables. 
 In recent years, the relativistic mean-field approximation
 has been successfully used for the analyses of $(e,e'p)$
 reactions
   in the 
 so-called relativistic distorted
 wave impulse approximation (RDWIA)  \cite{otros,Udi93,Udi95,Udi96,Granada}.
The polarization degrees of freedom
for  the electron and  the ejected nucleon
 have been included in this formalism years ago   \cite{vanorden}.
 In RDWIA, the
 nucleon  current
\begin{equation}
J^{\mu}_{N}(\omega,\vec{q})=\int\/\/ d\vec{p}\/ 
\bar{\psi}_F
(\vec{p}+\vec{q}) \hat{J}^\mu_N(\omega,\vec{q}\/) \psi_B(\vec{p})
\label{nucc} 
\end{equation}
is calculated with relativistic   $\psi_B$ and $\psi_F$ 
 wave functions for  initial bound
and  final outgoing  nucleons, respectively.  
$\hat{J}^\mu_N$ is the
relativistic nucleon current operator of $cc1$ or $cc2$ forms
\cite{deforest}.
 As bound state wave function,  Dirac-Hartree solutions from
relativistic Lagrangian with scalar and vector (S-V) meson terms
\cite{Horowitz}
 or
solutions of Dirac equation with phenomenological
 Woods-Saxon wells are customarily
used.
 The wave function 
 with asymptotic momentum 
$\vec{P}'$ for the outgoing proton is a  solution
 of the Dirac equation containing  S-V  optical potentials.
Recently a  relativistic calculation of $P_n$ following those
general lines has appeared
\cite{Sherif}. In reference  \cite{Sherif} it was found 
that the agreement with the data improved slightly compared to
the non relativistic analyses of reference \cite{Woo}. 
However, relativistic effects for this improvement 
remained unspecified in \cite{Sherif}.

Some of the differences between the relativistic and non relativistic
approaches  are independent of
the dynamics, having to do with the proper (relativistic) kinematics
being taken into account. Also, the non relativistic
operators are normally obtained from an expansion and truncation in
powers of $p/M$ and sometimes also of $q/M$ and $\omega/M$.
When the momenta and energy involved in the reaction are
of the order of the nucleon mass, as it may be the case
for $(e,e'\vec{p})$ reactions, 
one must be very careful with
the behavior of the expanded and truncated operator.
In reference \cite{Amaro} different non relativistic
expansions were studied and new
expressions that compared better with the unexpanded result were
deduced. 
In reference \cite{Ryckebush}  improved non relativistic
operators were used, particularly with the inclusion of the extra
spin-orbit term in the charge density operator as described
in reference \cite{Amaro}. This term proves to be necessary
to reproduce at least qualitatively 
\cite{Granada,Udi99}
the features seen in the $R_{TL}$ response and $TL$ asymmetry, 
$\displaystyle A_{TL}=
\frac{\sigma(\phi'=\pi)-\sigma(\phi'=0)}
{\sigma(\phi'=0)+\sigma(\phi'=\pi)},$
as measured in a recent TJNAF experiment at $Q^2\simeq 0.8$
(GeV/c)$^2$\cite{Gao}. 

 The non relativistic approach can be better compared to the relativistic
one thinking in terms of the direct Pauli reduction \cite{Udi95}.
Starting from a non relativistic formalism
based on bispinors $\chi$ solutions of a Schr\"odinger-like equation,
  one may at best
construct properly normalized four-spinors of the form
\begin{equation}
\psi_{nr}=\frac{1}{\sqrt{N}}
\left(\chi(\vec{p}),\frac{\vec{\sigma}\cdot\vec{p}}{E+M}\chi(\vec{p})
\right)
\label{eqx}
\end{equation}
to be introduced in Eq.~(\ref{nucc}) in order to
calculate a relativistic-like nucleon current matrix element. 
In this way the  relativistic kinematics is fully
taken into account and no expansions in $p/M$ are needed.
One further step to {\em relativize} the
calculations is done by rewriting  the Dirac equation
for the upper component
as a Schr\"odinger-like equation  and introducing  its  {\em non relativistic}
bispinor solution $\chi$   
 in Eq.~(\ref{eqx}). This `non relativistic'  bispinor is
phase-shift and
energy eigenvalue equivalent to the relativistic solution
\cite{Udi95,Cannata,Rawitscher,Kel96,Jin94}.
Comparing this solution of the Schr\"odinger-like 
equation  to the 
upper component of the fully relativistic wave functions, one finds
an additional factor (i) so that
the upper component of the full
Dirac solution is quenched in the nuclear interior
compared to the non relativistic solution \cite{Udi95,Cannata,Rawitscher,Jin94}.
This quenching can be associated to  
the Darwin factor\cite{Darwin} that appears from an extra term linear
in $\vec{p}$ that must be dealt with
to obtain  the Schr\"odinger-like equation and that is not present in the
usual 
non relativistic treatment.

One can then build a non relativistic formalism
based on the Schr\"odinger-like equation, with
central and spin-orbit potentials that are phase-shift equivalent to
the relativistic potentials,   incorporating {\em a
posteriori} 
the Darwin term in order to recover exactly the
same upper component as in  RDWIA and, by
means of Eq.~(\ref{eqx}), avoid the expansions
in $p/M$. This formalism would incorporate
all the kinematical and operator-related relativistic
effects, as well as the dynamical quenching of the
relativistic upper components due to the Darwin term.

This is done for instance by Kelly in several works
\cite{Kel97} though with an additional approximation,
the effective momentum approach (EMA) for the lower components.
This
 amounts to approximate the
 $\vec{\sigma}\cdot{\vec{p}}$ term that appears for the lower
components in Eq.~(\ref{eqx}) 
$(\displaystyle\chi(\vec{p})_{lower}=
\frac{\vec{\sigma}\cdot\vec{p}}{E+M}\chi(\vec{p})_{upper} 
)$
by $\vec{\sigma}\cdot{\vec{p}_{as}}$, with $\vec{p}_{as}$
the  momentum
corresponding to the asymptotic 
kinematics at the nucleon  vertex.
Results obtained within this approximation
both with relativistic and non relativistic
potentials
 were compared to
experiment in reference \cite{Woo}.

The differences between the calculations of \cite{Sherif}
and those presented in \cite{Woo} can  be either due to the
EMA procedure,  or to an additional dynamical relativistic effect different
from  the Darwin term, namely the  enhancement
of the lower components (ii):
The lower components of the fully relativistic solutions are enhanced 
at the nuclear interior due to the presence of negative energy
components\cite{Udi95,Cab98a,Ulrych}.  Solving the  Dirac equation with
scalar and vector potentials we see that the lower components
are related to the upper ones by
\begin{equation}
\chi(\vec{p})_{lower}=\frac{\vec{\sigma}
\cdot\vec{p}}{E+M+S-V}\chi(\vec{p})_{upper}.
\label{enhancement}
\end{equation}

Comparing with Eq.~(\ref{eqx}), we see that the
lower components are enhanced with respect to
the ones of free positive energy spinors by a factor $\displaystyle A^{-1}(r)=
\frac{E+M}{E+M+S(r)-V(r)}$ (we recall: $S<0, V>0$ and $A^{-1}(r)$ 
is $\simeq 2$ at the nuclear interior for the usual values of the
 potentials).
$A^{-1}(r)$ equals
 the inverse of the Darwin factor squared.
This enhancement of the lower components with regard
to free spinors has been sometimes referred as {\em spinor 
distortion} \cite{Kel97}.

As one can see
from Eqs.~(\ref{eqx}) and ~(\ref{enhancement}), for small values of the
 momentum $\vec{p}$ the lower 
components 
would play a minor role with respect to
the upper ones, due to the factor $\vec{p}/(E+M)$. In this
low-$p$ region ($p<300$ MeV), the enhancement of the lower components
is not important for  the $(e,e'p)$ cross sections\cite{Udi95}
 and the
most visible 
difference between RDWIA and  non relativistic (kinematically corrected) 
DWIA results is caused by the effect  
 mentioned above in point (i), namely the Darwin
term. Due to dynamical effects,
relativistic cross sections
at low-$p$ are smaller than the
non relativistic ones and  RDWIA--deduced spectroscopic factors 
from the low-$p$ data are 10\% to 15\% higher than the non relativistic
ones \cite{otros,Udi93}. 
With increasing $p$, however, the lower components cease to be small and
their enhancement, present in the fully relativistic wave functions
but not in Eq.~(\ref{eqx}) or in similar non relativistic expressions,
increases  the cross sections at $p>300$ $MeV/c$,
compared to the non relativistic ones. This 
improves sizeably the agreement with the data  of the RDWIA
$(e,e'p)$ cross sections \cite{Udi96,Granada}. In short, in regions
where the momenta of the bound and/or final nucleon are comparable
to the nucleon mass, the RDWIA  cross sections are
 {\em larger} than the non relativistic DWIA ones, in spite of the Darwin
factor that 
in these kinematical regions would play a minor role.
The more visible  dynamical effect in  high-$p$ regions 
would be the one mentioned in paragraph
(ii) 
that is, the enhancement
of the lower components. This enhancement is  crucial
to obtain good agreement\cite{Granada,Udi99} with the recent data for
$R_{TL}$ response and $A_{TL}$ in $^{16}O$ 
taken at
TJNAF for $Q^2\simeq 0.8$ $(GeV/c)^2$\cite{Udi99,Gao}.

To compare with non relativistic calculations,
one can project  the negative  energy sector out of the fully relativistic
solutions, thus  removing the enhancement of the lower components
described in paragraph (ii). 
More specifically, if the negative-energy components
are projected out, the nucleon
current is calculated  as
\begin{equation}
J^{\mu}_{proj}(\omega,\vec{q})=\int\/ d\vec{p}\/ \,
\bar{\psi}_F^{(+)}
(\vec{p}+\vec{q})
 \hat{J}^\mu(\omega,\vec{q})
\psi_B^{(+)}(\vec{p}),
\label{nuccproj}
\end{equation}
where $\psi_B^{(+)}$ ($\psi_F^{(+)}$) is the positive-energy component
of $\psi_B$ ($\psi_F$), {\em i.e.},
$\psi_B^{(+)}(\vec{p})=\Lambda_{(+)}(\vec{p}) \psi_B(\vec{p}),\,\,\,
\Lambda_{(+)}(\vec{p})=(M+\overline{\slash\!\!\! p})/2M ,$
with $\bar{p}_\mu=(\sqrt{\vec{p}^2+M^2},\vec{p}\,)$
(similarly for $\psi_F^{(+)}$).
That is, the matrix element of the current is computed
in a truncated space with only positive energy  spinors 
without enhancement of
the lower components. This truncation is inherent to all non relativistic
calculations.
The dynamical enhancement of the lower components  is contained
in the current of Eq.~(\ref{nucc}) but not in Eq.~(\ref{nuccproj}). Apart
from kinematical effects, the matrix elements obtained with the
prescription of Eq.~(\ref{nuccproj}) are equivalent to the ones computed in
non relativistic approaches based upon either the Foldy-Wouthuysen reduction
\cite{Giusti,McVoy} or the direct Pauli reduction \cite{Udi95,Poulis}.

The EMA approach  (or more properly the EMA-noSV one
 where, as we have said, no spinor distortion 
is considered\cite{Kel97}) also removes 
the enhancement of the lower components but it is not completely
equivalent to the exact projection method. Indeed, it  is 
equivalent to neglecting the $p$-dependence of the projection operators in
Eq.~(\ref{nuccproj}),  using instead the asymptotic values of the
momenta at the nucleon vertex. We can say that the EMA-noSV approach
computes the matrix element with spinors that have  the
same structure as the ones that enter in the scattering
of free nucleons, because it enforces the relationship 
between upper and lower components to be driven by the
asymptotic value of the  momenta at the nucleon vertex.  
The EMA-noSV calculation 
lacks   any `spinor distortion', exactly 
as in the  scattering of free nucleons. 
In particular, the Gordon transformation
is exact for the EMA-noSV approach and   $cc1$ and $cc2$
operators would lead to identical results within  EMA-noSV, if the same 
choices for the off-shell values of $\omega$, $E$, $E'$, $\vec{P}$ and
$\vec{P}'$ are made in both cases.

The projected results, on the other hand,
though lacking the large (around a factor of two) enhancement of the
lower components seen in the fully relativistic calculation, are based
on spinors whose upper/lower components verify Eq.~(\ref{eqx})
but with a wider value of
momenta than in  scattering from free
nucleons. Thus, even projected (non relativistic) results can
have a certain  degree of spinor distortion
compared to the free case due to the dispersion by the nuclear potentials.

We must keep in mind that both projected and EMA-noSV results   still
incorporate the dynamical quenching of the upper components
(Darwin term) and, if they are to be compared with
 non relativistic calculations,   care must be taken of the Darwin term
in the non relativistic result. 
Relativistic optical potentials normally give rise to increased absorption
and  stronger spin-orbit potentials. 
Due to this, it is expected that they
would also lead to a stronger induced normal polarization. 

We
want to emphasize that  the possible differences with the
former EMA-noSV analyses of Woo {\em et al.} \cite{Woo} are not due
to the use of a relativistic optical potential.
Both in references \cite{Woo}
and  \cite{Sherif}, results were presented with the same
potential EDAI-C that we use in the present work. 
 The Darwin term (leading to increased absorption)
was also included in a similar way to us. Thus, 
if there are differences between 
our results and those of \cite{Woo}, they must be due to
 relativistic effects additional to the Darwin term  and different
from the fact that the optical potential is relativistic or not.

\section{Results}
For the bound state wave functions we use the parameters of the set NL3
\cite{Ringnew} that reproduces adequately the known momentum
distributions at low-$p$ \cite{Steenhoven}.
We have also computed results with other bound state wave functions
and found the effects in $P_n$ to be very small up to $p\le 250$ MeV/c.
For the scattered proton wave function,  we use the energy-dependent
$A$-independent potentials 
derived  by Clark {\em et al.}  for $^{12}$C (EDAI-C) and
$^{16}$O (EDAI-O)\cite{Hama}. To study the sensitivity to different
optical potentials, we also compute results with the
energy-dependent $A$-dependent parametrization 1 (EDAD-1) of 
reference \cite{Hama}.

\subsection{Comparison with former results from Bates}

In Fig.~1,  $P_n$ is presented against the momentum of the recoiling
residual nucleus or {\em missing momentum} $p_m$,  related
to the momentum  of the  nucleon inside the 
target nucleus before being knocked-out
\cite{Boffi}.
 The results are computed for the kinematics of 
 reference \cite{Woo}, namely beam energy
of 580 MeV, kinetic energy of the final nucleon of 270 MeV,
$\left| \vec{q} \right|\simeq 760$~MeV/c, $\omega\simeq 290$ MeV in $q$-$\omega$ 
constant 
kinematics 
with $Q^2\simeq 0.5$ (GeV/c)$^2$. 
We use the Coulomb gauge in all the cases. We  included the Coulomb
distortion of the electron wave function and found its
effect in $P_n$ to be  small.
With solid (dotted) lines we present the fully relativistic results
obtained with the $cc1$ ($cc2$) operator. 
We
also show
results after projecting out the negative energy components 
 (short-dashed lines for $cc1$, long-dashed lines for $cc2$).

We see that for the fully relativistic results, the agreement with the
 data is excellent 
in both shells, except
perhaps for the highest $p_m$ and missing energy data points
in the $s_{1/2}$ shell, where the contribution from continuum
states not considered in the present work begins to be important.
Looking at the projected results, we see that the removal of the
negative energy components worsens the agreement with the data 
for both  the $cc1$ and $cc2$ operators. In all cases $P_n$ is smaller
(less positive or more negative) for the projected calculations.
We also see that
the Gordon ambiguities, {\em i.e.,} the differences between $cc1$
and $cc2$ results, are rather small. Compared to the theoretical
results of reference \cite{Woo}, the agreement with  experiment
is better for any of the curves presented in Fig.~1. 
This  cannot be  due only to the negative energy
components because the  effect of projection is rather  modest
 as the results
in Fig.~1 show.
To disentangle  the reasons for this difference
 in Fig.~2 we show (dotted lines) a calculation 
obtained with  the EDAI-C potential 
within the  EMA-noSV approach   with
the operator
$cc1$ (very similar
results  are obtained with the $cc2$ operator and are not shown here).
As a guidance, the solid line in Fig.~2 corresponds to the same one
of Fig.~1.  We see in Fig.~2
that for the EMA-noSV results the reduction of $P_n$ is  noticeable and
the agreement with the data  is  worse.
Our EMA-noSV results 
are  in the
line of the ones obtained 
 with the same optical potential EDAI-C
in reference \cite{Woo}.
We note that our $cc1$ and $cc2$ results are not identical  within
EMA-noSV due to the different off-shell prescription we use in each case
for the values of the kinematical quantities that enter in the evaluation
of the current matrix element.
Following reference \cite{deforest} for $cc2$ we have
\begin{equation}
\hat{\j}^\mu_{cc2}=F_1 \gamma^\mu+i \frac{\sigma^{\mu\nu}q_\nu}{2M}F_2,
\label{cc2}
\end{equation}
with $q^\nu=(\omega,\vec{q})$ at the electron vertex. For
$cc1$ we have
\begin{equation}
\hat{\j}^\mu_{cc1}=(F_1+F_2)\gamma^\mu - \frac{F_2}{2M}(P'+p_m)^\mu,
\label{cc1}
\end{equation}
with $p_m=(\overline{E},\vec{p}_m)$, $\vec{p}_m=\vec{P}'-\vec{q}$,
$E'=\sqrt{\vec{P}'^2+M^2}$ (the final nucleon is asymptotically on-shell)
and $\overline{E}=\sqrt{\vec{p}_m\,^2+M^2}$. Thus, $\omega$ used
in Eq.~(\ref{cc2}) is different from $\overline{\omega}=E'-\bar{E}$ as implied
in Eq.~({\ref{cc1}).
If we had used  $\overline{\omega}$ in $cc2$
instead of $\omega$, the results of our 
$cc1$ and $cc2$ EMA-noSV calculations would be identical. At present, there
is no definite prescription for handling this off-shell kinematical
ambiguity in $\omega$, $\overline{\omega}$ and other kinematical variables
to be used in the current operator \cite{Naus,Donnelly}. This ambiguity
arises   
because, contrary to the scattering of free nucleons,
  part of the energy and momentum of the exchanged photon is
transferred to the recoiling system instead of being completely
absorbed by the detected nucleon. 
We have
chosen the original prescription of de Forest of using $\overline{E}$ in
$cc1$ but $\omega$ in $cc2$. We find that, in this way, off-shell kinematics
ambiguity effects and Gordon ambiguity ones reinforce each other so that
the differences between the $cc1$ and $cc2$ results are enhanced by
our choice. The EMA-noSV results are free from Gordon ambiguities. This is
the reason
why $cc1$ and $cc2$ results obtained in such approach  are much
closer than the corresponding ones of the full RDWIA calculations.
 Kinematical ambiguities
will cause differences of up to 15\% between our $cc1$ and $cc2$
 EMA-noSV unpolarized cross sections
results for all the cases considered in the present  work. 
However, the effect of
these ambiguities in the  $P_n$ predictions 
is almost negligible (typically less than 1\%).
We have plotted only
the $cc1$ EMA-noSV  result for $P_n$. Our  $cc2$ EMA-noSV
result is almost identical to this curve.

Recently,
Kelly has incorporated the effect of spinor distortion
within his EMA approach by including 
the relativistic potentials in the lower component
of the  spinors as in Eq.~(\ref{enhancement}), while still 
substituting 
the $\vec{\sigma}\cdot\vec{p}$  term by $\vec{\sigma}
\cdot\vec{p}_{as}$. This is called EMA-SV approach in
reference \cite{Kel97} in contrast  to EMA-noSV. This procedure
reintroduces  the dynamical
enhancement of the lower components. For modest values 
of the momentum (up to around 275~MeV/c), where the role of the lower components
is less relevant, this approximation
goes close\cite{Kellyprivate} to the results obtained with
the exact treatment 
that we do in the present work. It tends, however, to minimize 
the effect of enhancement of the lower components beyond the value of
$p_m$ mentioned above
and, as  the momentum of
the ejected nucleon is normally well above  275 MeV/c, it 
underestimates the effect of spinor distortion for the ejected nucleon.
We conclude that an important reason for the differences of our
full RDWIA results with those
 of \cite{Woo},  using the same optical potential
EDAI-C, is  the EMA-noSV approach employed in reference \cite{Woo}.

For the purpose of comparison, we present 
  (dash-dotted line in
Fig.~2)
results obtained 
with 
the same bound state wave functions
as for the other curves, the full RDWIA  approach with
the $cc1$ operator but
a different optical potential, namely the EDAD-1 of ref.\cite{Hama}.
We emphasize that the EDAI-C potential should be
a more suitable choice than EDAD-1 for $^{12}C$ because  it describes
better the elastic proton scattering data for this particular
nucleus. The
effect  of using EDAD-1 instead of EDAI-C is sizeable. EDAD-1
yields a
larger $P_n$ for 
the $p_{3/2}$ shell, worsening the  agreement with the data
 and  a smaller $P_n$ for the $s_{1/2}$ shell,  with
no significant worse (or better) agreement with the data in this shell.
We observed that both EDAI-C and EDAD-1 produce almost identical
unpolarized cross sections or unpolarized response. However, the
$P_n$ values they produce are noticeably different. 
This shows the sensitivity of
$P_n$ to details of the FSI.

The comparison between the results of both potentials
follows the same general trend as  shown in
reference \cite{Sherif}. In reference  \cite{Sherif},
$P_n$ changes in the same relative
direction in going from EDAI-C to EDAD-1
as in our calculation but,  contrary to our case, the differences
between the EDAI-C and EDAD-1 are larger for the $1s_{1/2}$ shell
than for the $1p_{3/2}$ one and in this latter shell,  their
results  seem to agree better with experiment for the EDAD-1 potential
than 
for the EDAI-C. We  find, in general, that our
results with EDAI-C  agree better with the data.
 These minor discrepancies
with reference \cite{Sherif} should be
traced back to the different bound state wave functions
and possibly other parts of the formalism (current operator, Coulomb
distortion) employed in both cases.

In Fig.~2 we also show 
the EMA-noSV results 	of reference \cite{Woo} 
with the EEI potential
(dashed line). In reference \cite{Woo}  
a more
positive $P_n$ for the $p_{3/2}$ shell and  better
agreement with the data was obtained
 with the EEI potential than with the EDAI-C one.
For the $1s_{1/2}$ shell, however,
a more negative $P_n$ and worse agreement 
with the data was found.

The EMA-noSV-EEI calculation of reference \cite{Woo} 
underestimated
the data by about  10\%.
The EEI is a non relativistic  optical potential obtained after
 folding a density-dependent empirical effective interaction
with the nuclear density\cite{EEI}. 
The interaction is fitted to proton-nucleus
elastic {\em and inelastic} scattering data for several states
of several targets simultaneously. The relativistic optical potentials,
on the other hand, are fitted only to elastic scattering data.
It is well known that elastic data can only constrain the
asymptotic part of the potentials. Given the
fact that the nucleus is almost transparent to electrons
(compared to nucleons), phase-shift equivalent
potentials  that differ only in the nuclear interior 
would produce  the same good fits to elastic proton
scattering observables, 
leading however to different electron scattering results.
The EEI approach solves this ambiguity by 
phenomenologically constraining the potentials in the
nuclear interior by means of simultaneous fits to inelastic data.
In the relativistic case, on the other hand, the shape of the potentials
at the nuclear interior is assumed to be of simple Woods-Saxon  +
surface terms, not very different from what one finds
in the relativistic mean field approximation.
Thus, the fact that the relativistic model gives  a very fair
account of $(e,e'\vec{p})$ observables such as $P_n$
cannot be attributed merely to the
incorporation of the right phenomenology, as it could be the
case with the EEI potentials, but 
to a merit of the model itself.


\subsection{Predictions for Mainz and TJNAF in parallel kinematics}
In a recent work \cite{Granada,Udi99,Cab98b}, it has been
 shown that the
dynamical 
enhancement of the 
lower components shows up differently in the $j=l-1/2$ and
$j=l+1/2$ spin-orbit partners, specially for the $R_{TL}$ 
response and the $A_{TL}$ asymmetry. We remind that for $^{12}C$ the two
shells 
studied  correspond to $j=l+1/2$ spin-orbit partners.
New
sets of induced polarization $P_n$ are being obtained
at TJNAF\cite{Glasshauser} in $^{16}O$ at   
a more relativistic kinematics,
namely beam energy of 2450 MeV, kinetic energy of the
ejected nucleon of about 420 MeV, and $Q^2\simeq 0.8$~(GeV/c)$^2$.
There is also a proposal at Mainz\cite{MAMI} to do
similar measurements at a  smaller value of the
kinetic energy of the ejected nucleon, namely 200 MeV. In
what follows, we analyze
 whether 
these experiments may provide  
 signatures of relativistic dynamics in $P_n$ similar to the ones 
found in $R_{TL}$ and $A_{TL}$.

In parallel kinematics only two responses, $R_L$ and $R_T$, contribute
 to the unpolarized cross section and just one, $R_{TL}^n$, to $P_n$,
 so that
for this kinematics the analyses are simplified.
 We present in Fig.~3 
fully RDWIA results with $cc1$ (solid line), $cc2$ (dotted line),
projected results (short-dashed lines for $cc1$ and long-dashed lines for $cc2$)
and EMA-noSV 
results with the $cc1$ operator (dash-dotted line)
 for the two $p$-shell spin-orbit partners
of $^{16}O$, plus the deep $s$-shell, in  
 parallel kinematics and with beam energy and transfer energy
suitable for Mainz\cite{MAMI}
 (beam energy of 855 MeV, kinetic energy of the 
ejected nucleon of 200 MeV).

For  the three shells we see
 the opposite pattern   to the 
 one depicted in Fig.~1: the removal of the
negative energy components drives here $P_n$ towards higher values,
and even more so does using the EMA-noSV approach.
For  the kinematics of Fig.~3, the effect of
projection and Gordon ambiguities is much larger 
than it was in Fig.~1.

A very  characteristic feature is seen in the $s_{1/2}$ shell for 
this case of parallel
kinematics:
A zero value of $P_n$ is predicted within the
EMA-noSV approach. 
A small value of $P_n$ is obtained by the
projected calculations, 
while the full RDWIA approach yields a relatively large 
(in absolute value) 
 $P_n$ due to spinor distortions. The choice $cc1$, that emphasizes the effect of
the enhancement of the lower components \cite{Cab98a}, yields the 
largest prediction for  $P_n$ in absolute value.
Should the experiments at TJNAF or Mainz provide us with $P_n$
values with equal or smaller uncertainty that the ones already
measured at BATES, it will undoubtedly  disentangle the role
played in $P_n$ by the enhancement of the lower components. 

The responses involved in the evaluation of $P_n$ for
this case are displayed in Fig~4. 
The only nonzero contribution to $P_n$
comes from $R_{TL}^n$  and 
the link between the responses
shown in the bottom  panel of Fig.~4 and the results for
 $P_n$ of Fig.~3 is straightforward.
As it could be deduced from the values of
 $P_n$ displayed in Fig.~3,
$R_{TL}^n$ for the $s_{1/2}$ shell is zero within EMA-noSV, 
it is very small for the projected results and
 reaches the largest absolute 
value for the full RDWIA $cc1$ calculation.
The results for the  $p_{3/2}$ shell follow the same trend as
shown for the $s_{1/2}$ shell, only that here the more complex
spin-orbit structure of the bound state causes 
a nonzero value of $P_n$ even for the EMA-noSV results.
The projected and EMA-noSV results display  small
 (in absolute value)
 predictions for
 $R_{TL}^n$.
 The $cc2$ RDWIA prediction 
exhibits  larger  $R_{TL}^n$, 
while the full $cc1$ result yields
 the largest
value of $R_{TL}^n$.
This gradation of $R_{TL}^n$ is similar to what one finds  generally 
for the unpolarized $R_{TL}$ (in  $q$-$\omega$ constant kinematics): 
As the $cc1$  operator 
 enhances the role of the negative energy components \cite{Cab98a}
with regards to other choices of the operator, it  produces the
largest value of $R_{TL}$. Thus, at least for the $j=l+1/2$ spin
orbit partner, we observe the same behavior for $R_{TL}^n$ and
$R_{TL}$ with regards to the effect of negative energy components.

On the other hand, we  find that  projection and 
Gordon ambiguities effects show up  differently 
for the $p_{1/2}$ shell. For this case
$R_L$,
$R_T$ and $R_{TL}^n$  are shown in the leftermost panel of 
Fig.~4 at  the same kinematics
of Fig.~3. 
While for   $R_{TL}^n$ in the $j=l+1/2$ shells
the larger was the role given to the negative energy 
components the 
larger (in absolute value) $R_{TL}^n$ response was obtained,
for the $p_{1/2}$ shell ($j=l-1/2$)
one sees the opposite behavior: the full $cc1$ calculation 
yields the smallest $R_{TL}^n$ while the EMA-noSV prediction
displays the largest one.
This is at variance with the behavior observed for the 
unpolarized $R_{TL}$ response \cite{Cab98b} 
and  
indicates an interference between positive and negative energy
component contributions to $R_{TL}^n$. This interference is
constructive for the $j=l+1/2$ shells so that the calculations with
 large contribution from negative energy components 
yield a large $R_{TL}^n$, while it is largely destructive for
the $j=l-1/2$ shells for which  
 large  effects of   negative energy  components
translate into small values of $R_{TL}^n$.

In reference~\cite{coupled} results  were presented for the
EMA-noSV  case
within the
IA and also in a calculation beyond IA that 
included channel coupling to
several configurations in the final state. Our EMA-noSV 
result of Fig.~4 and 
the 
one shown in Fig.~14 of reference~\cite{coupled} are very similar, 
with small
differences due to the different wave functions and optical potentials.
The most interesting outcome of this comparison is that the effect of 
spinor distortion  increases $R_{TL}^n$, in particular for the
case of the $s_{1/2}$ shell that would have a zero value
 without spinor distortion
within  IA. Due to channel coupling (CC), a nonzero 
$R_{TL}^n$ for this shell
was obtained in reference~\cite{coupled}. 
The effect of spinor distortion,
however, is at least twice to four times 
(depending on whether one considers the RDWIA  $cc2$ or $cc1$ result) 
larger than the one of CC shown 
in reference~\cite{coupled}.
We conclude that
coupled channel contributions would not  
 mask the  large negative value of $P_n$ caused 
by spinor distortion. 
%
 The $R_{TL}^n$ response in this $s_{1/2}$ shell 
is  sensitive 
to Gordon ambiguities
and   overall constitutes 
a very clear signature for the presence of  
negative
energy components in the nucleon wave function.

For the other shells,  the effects of CC shown in
reference \cite{coupled} were small at moderate 
values of $p_m$
for the cases we studied in the present work and the IA results shown 
here should not 
 change much if CC effects were considered.

Still in parallel kinematics but with a larger value of $Q^2$ that
is suitable at TJNAF,
we have obtained very similar results to the ones just
 presented.
 We plot in Fig.~5 only the results for $P_n$.

\subsection{$q$-$\omega$ constant kinematics}

The experiment of Bates was performed in $q$-$\omega$ constant
 kinematics
and in the same side of $\vec{q}$ for the scattered proton, 
$\phi'=\pi$, which   corresponds to $p_m>0$ in our figures.
The analysis of this case is more complicated because all the eight
 responses
of Eq.~(\ref{responses}),
in combination with the factors shown in Table~1,
 contribute to the cross section
and $P_n$. 
In Fig.~6 and 7 we present the responses for the Bates
results depicted in Fig.~1. 
The effect of spinor distortion and Gordon ambiguities in the
$R_L$, $R_T$, $R_{TL}$ and $R_{TT}$ has been studied
previously in the context of RPWIA \cite{Cab98a,Cab98b}. It
was found there that for the $j=l+1/2$ partners, as it is
the case of the two shells in $^{12}C$, the differences are
relatively small, at least for the 'large' responses $R_L$ and
$R_T$.

For the normal responses the situation is  less clear.
One must remember that unpolarized and normal responses 
share the same structure in terms of components of the
hadronic current, differing only in the  signs
with which the  contribution for every value of
ejected  nucleon spin projection upon the
normal direction enters
into the unpolarized or normal responses\cite{donRask,vanorden}.
Thus, large  unpolarized responses usually
come from constructive interference of the  two
spin contributions and are associated with 
a correspondingly small normal response coming from   destructive
 interference.
The converse is also true: small unpolarized responses have
a correspondingly large polarized normal response \cite{vanorden}.
If there were no FSI, both
normal projection contributions (spin up and spin down) would
be identical, all the responses shown in Fig.~7 would
vanish and no normal polarization would be observed.

 Taking into account the value of the kinematical factors
in front of each response (see Table I),
the main 
 contributions to $P_n$ 
 for the $s_{1/2}$
shell
comes from $R^n_T$ and $R^n_{TL}$
 responses.
In the case of the full RDWIA
results,
the $R^n_{TL}$ response is responsible for
 most of the net $P_n$. 
 Due to this, $P_n$ will  change sign with $p_m$ because of the 
$\cos\phi'$ factor in Eq.~(\ref{responses}).
  We also see that $R_{TL}^n$ is larger (in absolute
value) for the calculation with larger
effect of the negative energy components,
{\em i.e.,} the full $cc1$ result. In Fig.~6 we can see the same feature in 
the unpolarized $R_{TL}$ response. This characteristic has been 
explained before
\cite{Udi99,Cab98a,Cab98b}. 
 For the $p$ shell,  the largest contribution comes from  $R_{T}^{n}$
that shows little dependence on spinor distortion.
This is why
the effect of negative energy components is small for this shell.

We plot also $P_n$ and responses (Figs. 8, 9 and 10) for
a kinematics suitable at TJNAF (namely $|\vec{q}|=1000$ MeV/c, 
$\omega=445$ MeV and energy of the beam $\epsilon=2445$ MeV). 
Apart from what
has  been already said, we find that
for the $p_{3/2}$ shell and $p_m<0$ there are  small Gordon
ambiguities and a very clear separation of the fully RDWIA
results from the projected or
EMA-noSV ones is seen. Therefore,
this is  a good region to look for the effects of 
spinor distortion.
We can explain  this better by looking at the results
 in the second column of 
Fig.~10: There, all  calculations lie very close 
except for  $R_{TL}^n$. 
In the $p_{3/2}$ shell  the projected and
EMA-noSV curves group together, while both fully relativistic 
calculations  
clearly deviate from the others. 
Going back to $P_n$ in the second
panel of Fig.~8, we observe these differences only in the region 
$p_m<0$, due to the different  
sign with which  the $\cos\phi' R_{TL}^n$ term contributes in
the $p_m<0$ and $p_m>0$ regions. 
This behavior is characteristic of the kinematics chosen at TJNAF.
Indeed, as we can see in Fig.~11, at
different kinematics conditions such as the ones suitable at Mainz, ({\em i.e.,}
$q$-$\omega$ constant kinematics with $|\vec{q}|=648$ MeV/c, $\omega=215$ MeV
and $\epsilon=855$ MeV),  there is not such a clear
separation of the fully relativistic curves from the others in the $p_{3/2}$
shell for $p_m<0$ as the one found  for the TJNAF kinematics.

Another interesting feature that was already found  in parallel kinematics 
is that, for the $j=l+1/2$ shells,
$R^n_{TL}$ has larger values when the calculation emphasizes the role
of negative energy components while the converse is seen for the
$p_{1/2}$ ($j=l-1/2$) shell. 

For the two $p$ shells at the kinematics of TJNAF and
Mainz,  the largest contribution to $P_n$
would come from  $V_{TT} R_{TT}^n$.
For the $p_{3/2}$ shell in the $p_m<0$ region, however,
 this contribution is 
canceled to a large
extent 
 by the $V_{TL} R^n_{TL}$ one. This
explains why $P_n$ is mainly negative for $p_m<0$ for the $p_{3/2}$
shell. 
 On the other hand, the $v_{TL} R_{TL}^n$ contribution  
is less important for the $p_{1/2}$ shell 
and practically does not influence the
total polarization. 
Therefore $P_n$ for this shell is negative 
irrespectively of the sign of $p_m$. 
 
A serious concern is the issue of current conservation. The use of
an optical potential breaks Gauge invariance in DWIA. We
estimated the uncertainty associated to the choice of Gauge by
comparing the results we show in the present work with the ones
obtained in the Landau Gauge. We find that the fully relativistic
results in perpendicular kinematics at the highest value
of $Q^2$ are the less sensitive ones to this procedure. 
On the other hand, the unpolarized cross sections can differ
by as much as 50\% in parallel kinematics for large values of $|p_m|$.
However, $P_n$ is much less sensitive to the choice of Gauge
than unpolarized cross sections: 
For the fully RDWIA results of the present work,
using the Landau or Coulomb Gauge produces $P_n$ results 
within 5\%.
Gauge ambiguity is much 
less important than the one due to the $cc1$ or $cc2$  choice.

\section{Conclusions}
We have found that the relativistic dynamical
effect mentioned in paragraph (ii), the enhancement of
the lower components, increases noticeably  $P_n$
with respect to both the projected and, more sizeably, the EMA-noSV
results,
driving  the fully RDWIA results for $P_n$ into
excellent agreement with the data of reference \cite{Woo}.
For the kinematics of the TJNAF 89-033\cite{Glasshauser}  
and Mainz \cite{MAMI} experiments
we find the differences between the RDWIA and
projected results to be  important. 

$P_n$ proves to be
very sensitive to the choice of optical potential, allowing 
this observable to be used to constrain the theoretical
model for FSI so that these effects can be included with confidence
when making predictions for other observables much
less sensitive to the choice of FSI, such as
the polarization transfer observables $P'_x$ and $P'_z$ \cite{Kel97}.

Previous explorations of the role of meson exchange
currents (MEC) for Bates, based upon a non relativistic picture,
 showed very little effect in $P_n$ at  moderate
$p_m$ \cite{Ryckebush}. MEC  are expected to play   an even
 minor role for higher $Q^2$ 
at  quasi-elastic kinematics ($x\simeq 1$)\cite{Ryckebush} 
and its inclusion
will not  modify substantially the predictions for $P_n$ presented
in the present work. The same can be said of coupled channel effects
analyzed 
within the EMA-noSV approach
in reference~\cite{coupled}. 
However, Gordon and kinematical off-shellness ambiguities are 
 large
for  high $Q^2$ experiments.  We have looked for
kinematical  regions 
where these ambiguities are minimized.
In parallel kinematics, we conclude that the $p_m<0$ region
($|\vec{q}|>|\vec{P'}|$)
is optimal because it displays a minimum
effect of Gordon ambiguities
while   a high sensitivity of the calculations
to the presence of negative energy components is found.
In $q$-$\omega$ constant kinematics 
the same favorable situation is seen again  
for $p_m<0$ ($\phi'=0$)  but only 
 for the $p_{3/2}$ shell at
 $|\vec{q}|=1$ GeV/c,  adequate to TJNAF.

In parallel kinematics we found very clear 
signatures for negative energy components in the wave functions,
 that cause 
  $P_n$ to be  driven towards more negative values 
with respect to the non relativistic prediction,
particularly for the $s_{1/2}$ and $p_{3/2}$  ($j=l+1/2$)
shells. This feature
should remain even in the presence of MEC and CC. 

In $q$-$\omega$ constant kinematics the effect of the negative
components is manifested as
an increase (decrease) of $P_n$ for $p_m>0$ ($p_m<0$) 
of the relativistic predictions with regards to the
non relativistic ones.

Finally, we hope that future experiments will shed light
on the theoretical
uncertainties
that are still present in the  calculations such as which current
operator should be used 
and will help  to disentangle the role played
by the negative energy components of the
wave functions.

This work was partially supported under Contracts  
 PB/96-0604 (DGES, Spain) and
PR156/97 (Complutense University, Spain). J.M.U.
thanks E. Moya de Guerra and J.A. Caballero for many useful
comments. We also thank J.J. Kelly for providing us with files
with his calculations and interesting comments.
J.R.V. acknowledges financial support from 
the Consejer\'{\i}a de Educaci\'on y Cultura de la
 Comunidad de Madrid.

%
%
\begin{table}
\begin{tabular}{cccc}
              &   TJNAF   &      MAMI   &      BATES  \\ \hline
$\theta_e$    &   23.4$^o$&     48.8$^o$& 118.8$^o$    \\ 
 $Q^2$ $(GeV/c)^2$    &  0.8    &      0.4    &      0.5    \\ 
 $\omega$  (MeV) &  445     &     215     &   292          \\
                &           &             &            \\
      $V_L$   &     0.643  &      0.792    &    0.727       \\   
      $V_T$   &     0.444  &      0.651    &    3.283     \\
      $V_{TL}$  &     0.737  &      0.932    &    1.642    \\
      $V_{TT}$  &     0.401  &      0.445    &    0.426
\end{tabular}
\caption{Approximated values of the
kinematical variables and the factors in Eq.~(\protect\ref{responses}) 
for the $q$-$\omega$ constant experimental
setups discussed in the present  work.}
\end{table}

\begin{figure}
\begin{center}                                                                
\leavevmode
\mbox{\epsfig{file=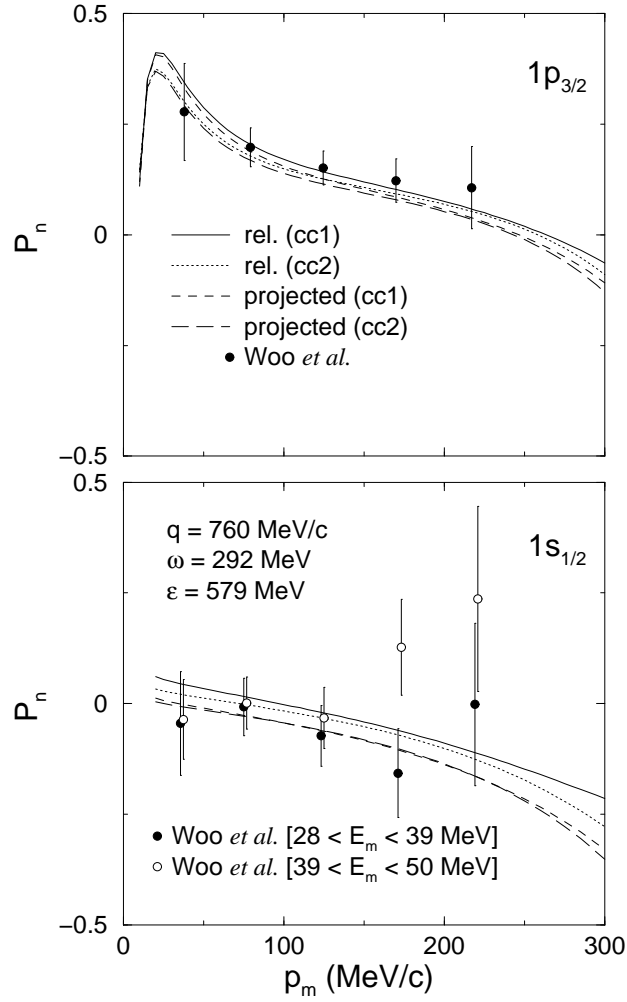,width=0.5\textwidth}}
\end{center}
\caption{
$P_n$  from $^{12}$C for
the $1p_{3/2}$ (upper panel) and $1s_{1/2}$ (lower panel)
orbits, versus missing momentum $p_m$ in MeV/c. Results shown 
correspond to a fully relativistic
calculation with the $cc1$ (solid) and $cc2$ (dotted)  operators.
 Also shown are the results after 
projecting the bound and scattered proton wave functions over positive-energy 
states (short-dashed and long-dashed lines respectively). Data points are from
\protect\cite{Woo}.} 
\end{figure}

\begin{figure}[t]
\begin{center}                                                                
\leavevmode
\mbox{\epsfig{file=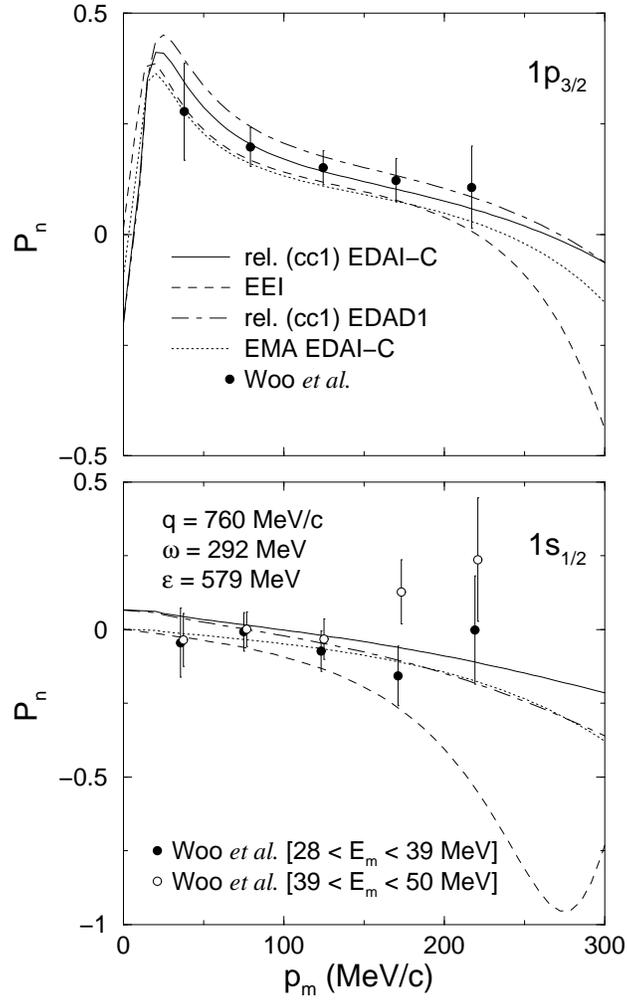,width=0.5\textwidth}}
\end{center}
\caption{
$P_n$  from $^{12}$C for
the $1p_{3/2}$ (upper panel) and $1s_{1/2}$ (lower panel)
orbits, versus missing momentum $p_m$ in MeV/c. Results shown 
correspond to a fully relativistic
calculation with the $cc1$  and the EDAI-C (solid),
and the EDAD1 (dash-dotted) potentials.
 Also shown are the  EMA results 
(dotted  line) for the EDAI-C case.
 Former EMA-EEI results (dashed line) and
data points are from  \protect\cite{Woo}.}
\end{figure}

%
%
\begin{figure}[t]
\begin{center}
\leavevmode
\mbox{\epsfig{file=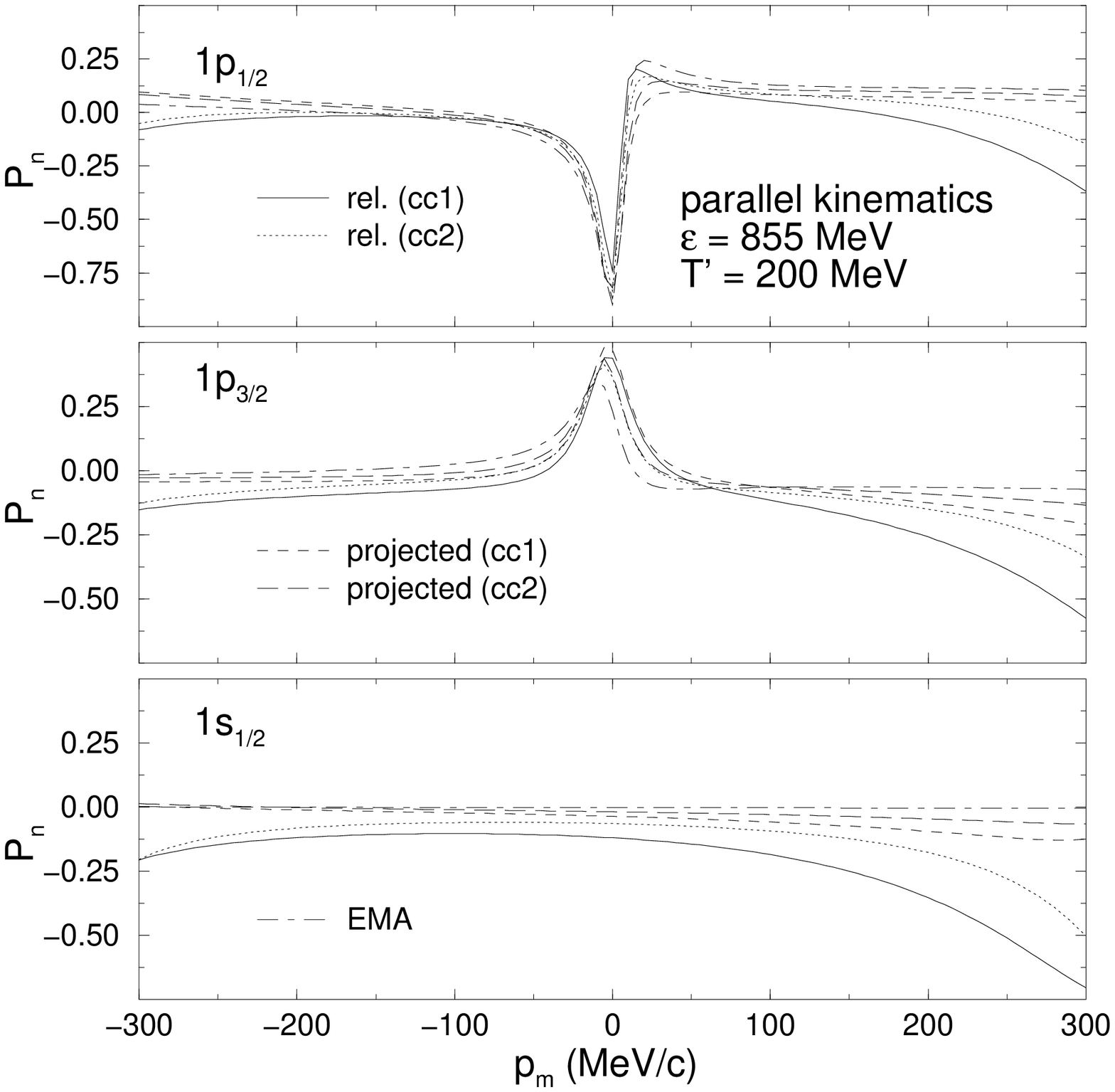,width=0.6\textwidth}}
\end{center}
\caption{$P_n$  from $^{16}$O for
the $1p_{1/2}$ (upper panel), $1p_{3/2}$ (mid panel) and
$1s_{1/2}$ (lower panel) 
orbits, versus missing momentum $p_m$ in MeV/c. Results shown 
correspond to a fully relativistic
calculation with the $cc1$ (solid line) and $cc2$ (dotted line)  operators.
 Also shown are the projected
results (short and long-dashed lines) and the EMA-$cc1$ ones
(dash-dotted line). 
Results in parallel kinematics corresponding to ref. \protect\cite{MAMI}
and the EDAI-O potential is used.}
\end{figure}

%
%
\begin{figure}[t]
\begin{center}
\leavevmode
\mbox{\epsfig{file=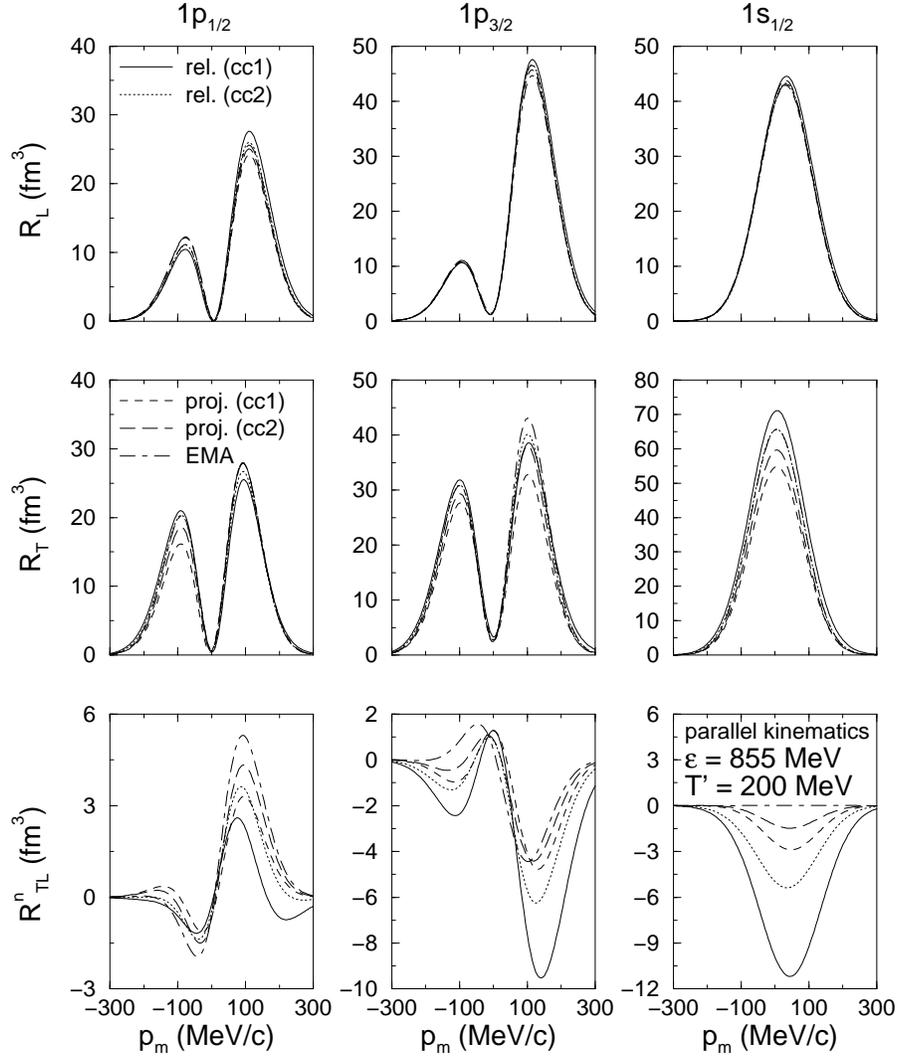,width=0.7\textwidth}}
\end{center}
\caption{Response functions corresponding to the kinematics of  Fig.~3.
}
\end{figure}

%
%
\begin{figure}[t]
\begin{center}
\leavevmode
\mbox{\epsfig{file=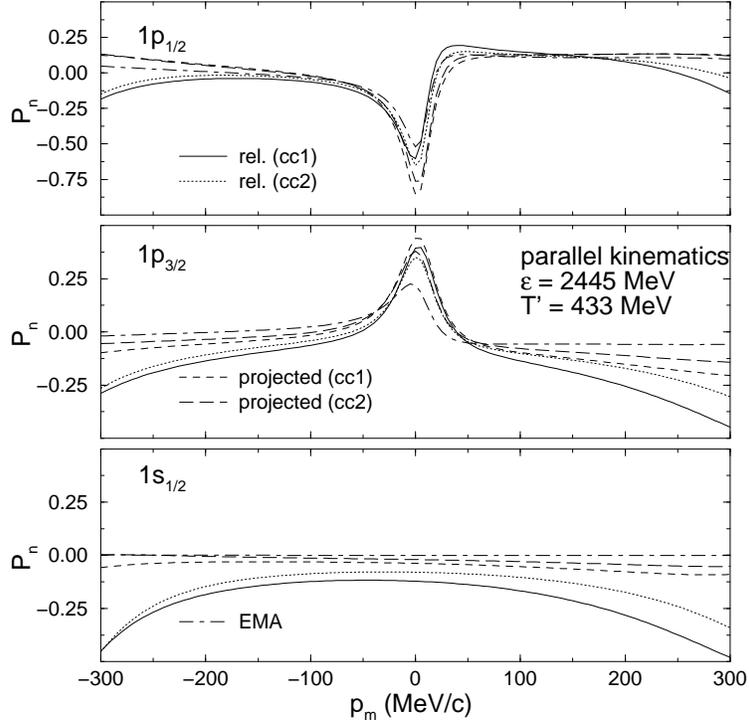,width=0.6\textwidth}}
\end{center}
\caption{$P_n$  from $^{16}$O for
the $1p_{1/2}$ (upper panel), $1p_{3/2}$ (mid panel) and
$1s_{1/2}$ (lower panel) 
orbits, versus missing momentum $p_m$ in MeV/c.
 Results shown 
correspond to a fully relativistic
calculation with the $cc1$ (solid line) and $cc2$ (dotted line)  operators.
 Also shown are the projected
results (short and long dashed lines) and
 the EMA(noSV)-$cc1$ ones
(dash-dotted line). 
Parallel kinematics suitable at TJNAF \protect\cite{Glasshauser}
and the EDAI-O potential is used.}
\end{figure}

%
%
\begin{figure}[t]
\begin{center}
\leavevmode
\mbox{\epsfig{file=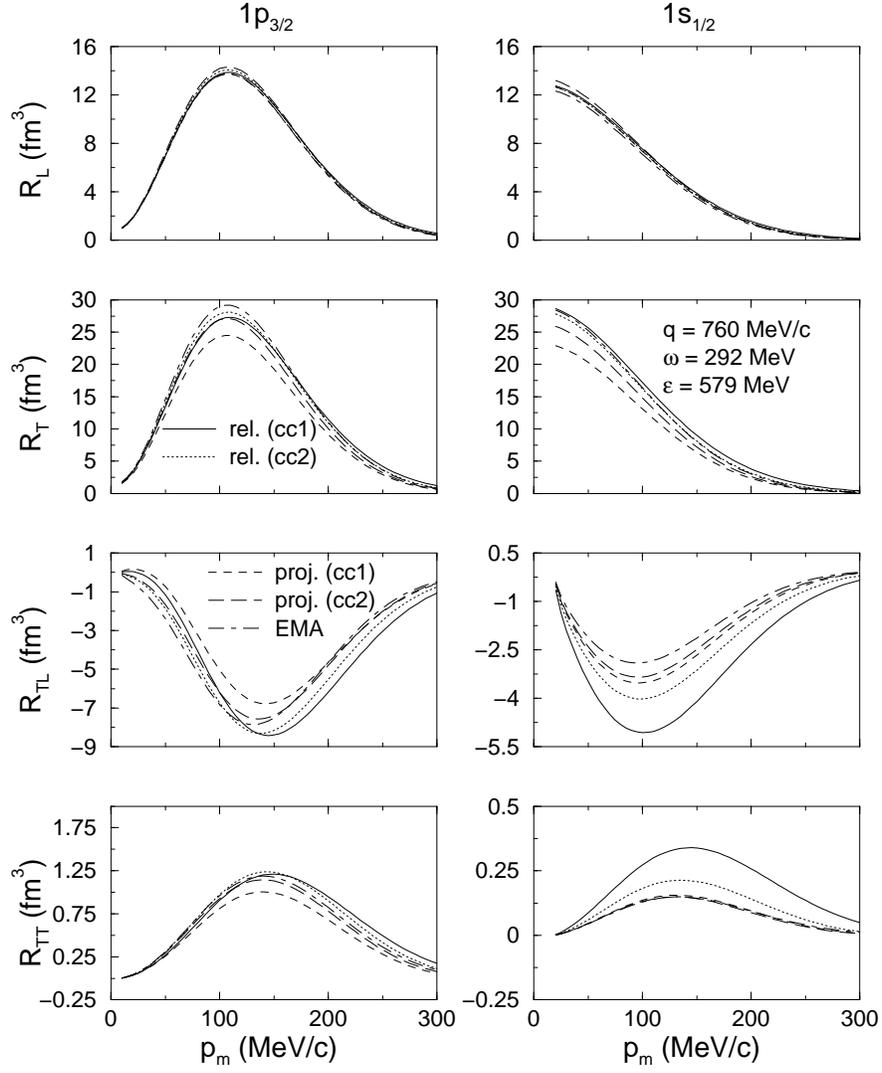,width=0.7\textwidth}}
\end{center}
\caption{Unpolarized responses for the kinematics of Fig.~1.}
\end{figure}

\begin{figure}[t]
\begin{center}
\leavevmode
\mbox{\epsfig{file=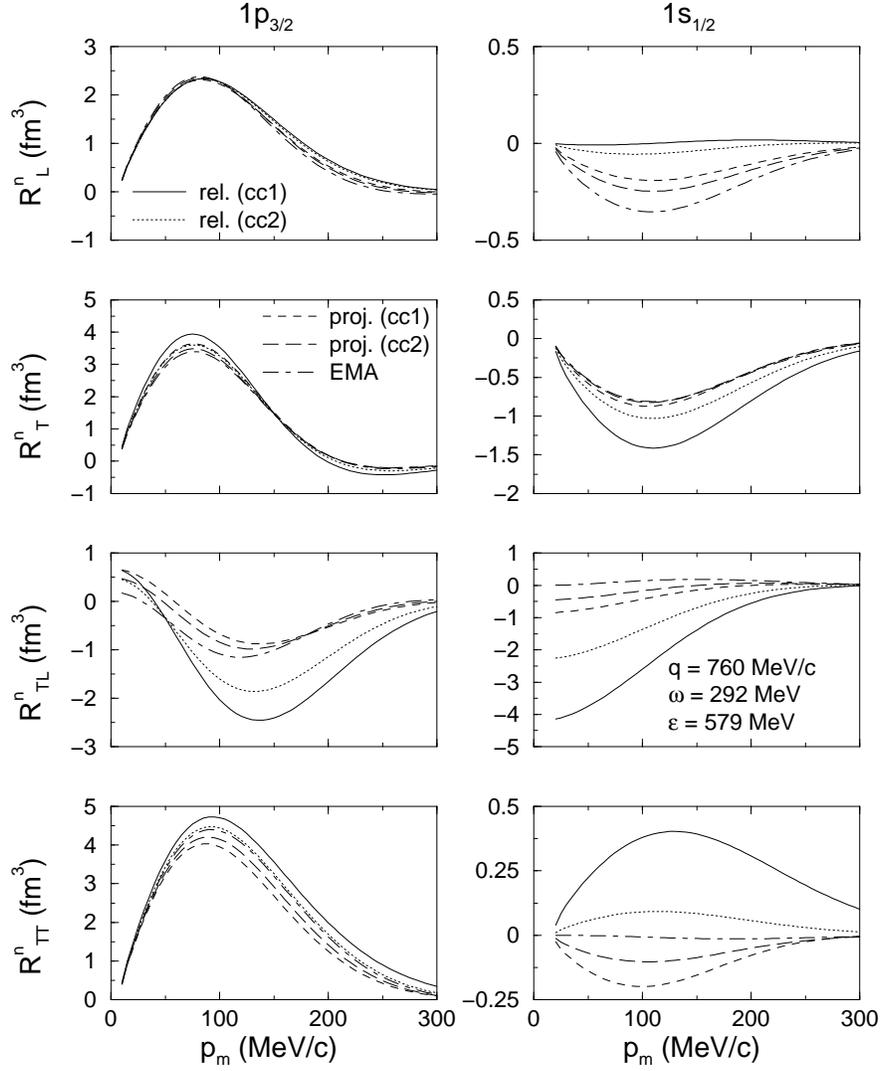,width=0.7\textwidth}}
\end{center}
\caption{Normal responses for the kinematics of Fig.~1.}
\end{figure}

%
%
\begin{figure}[t]
\begin{center}
\leavevmode
\mbox{\epsfig{file=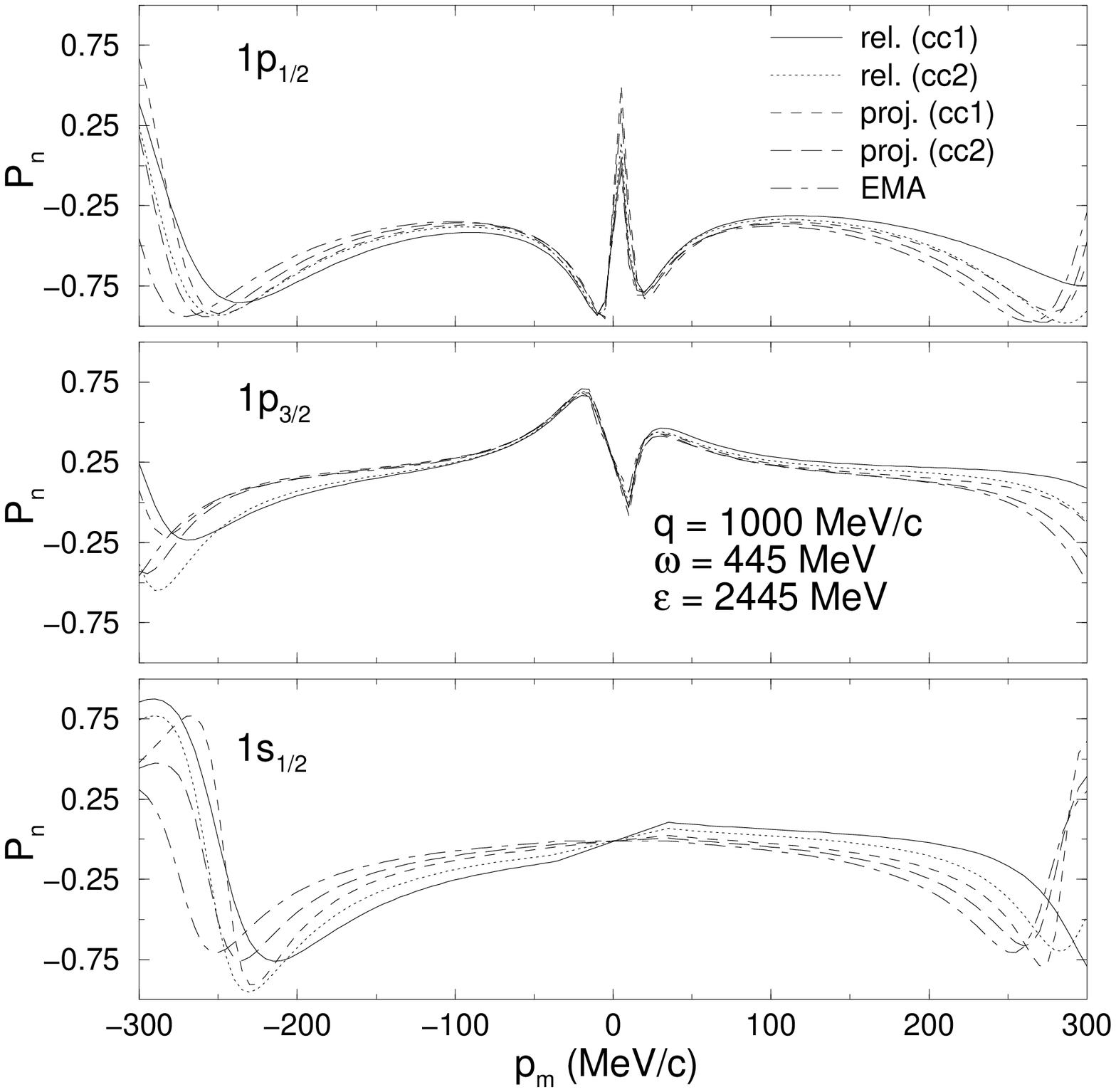,width=0.6\textwidth}}
\end{center}
\caption{$P_n$  from $^{16}$O for
the $1p_{1/2}$ (upper panel), $1p_{3/2}$ (mid panel) and
$1s_{1/2}$ (lower panel) 
orbits, versus missing momentum $p_m$ in MeV/c.
 Results shown 
correspond to a fully relativistic
calculation with the $cc1$ (solid line) and $cc2$ (dotted line)  operators.
 Also shown are the projected
results (short and long dashed lines) and the EMA(noSV)-$cc1$ ones
(dash-dotted line).
 Results in $q$-$\omega$ constant kinematics corresponding 
to ref. \protect\cite{Glasshauser}
and the EDAI-O potential is used.}
\end{figure}

%
%
%
\begin{figure}[t]
\begin{center}
\leavevmode
\mbox{\epsfig{file=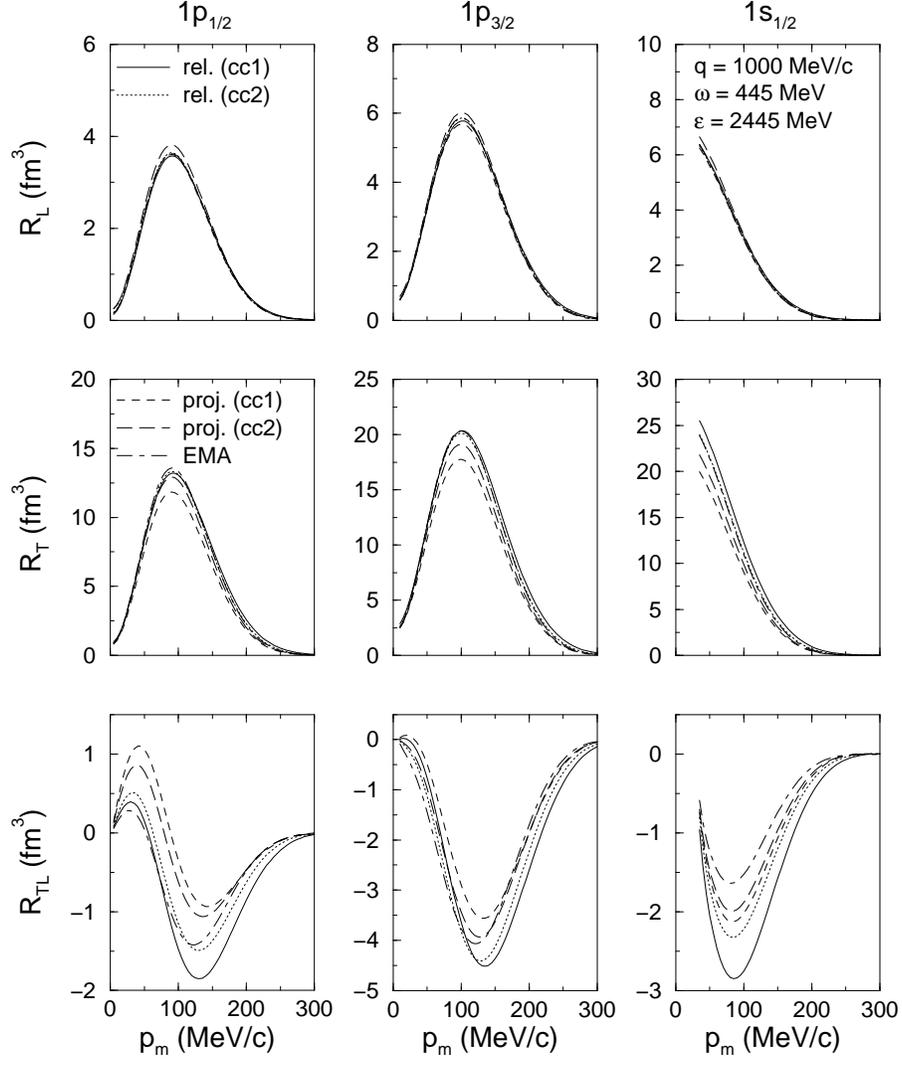,width=0.7\textwidth}}
\end{center}
\caption{Unpolarized responses for the kinematics of Figure 8. 
The response
$R_{TT}$ is not plotted, because it is much smaller than the  ones shown.}
\end{figure}

\begin{figure}
\begin{center}
\leavevmode
\mbox{\epsfig{file=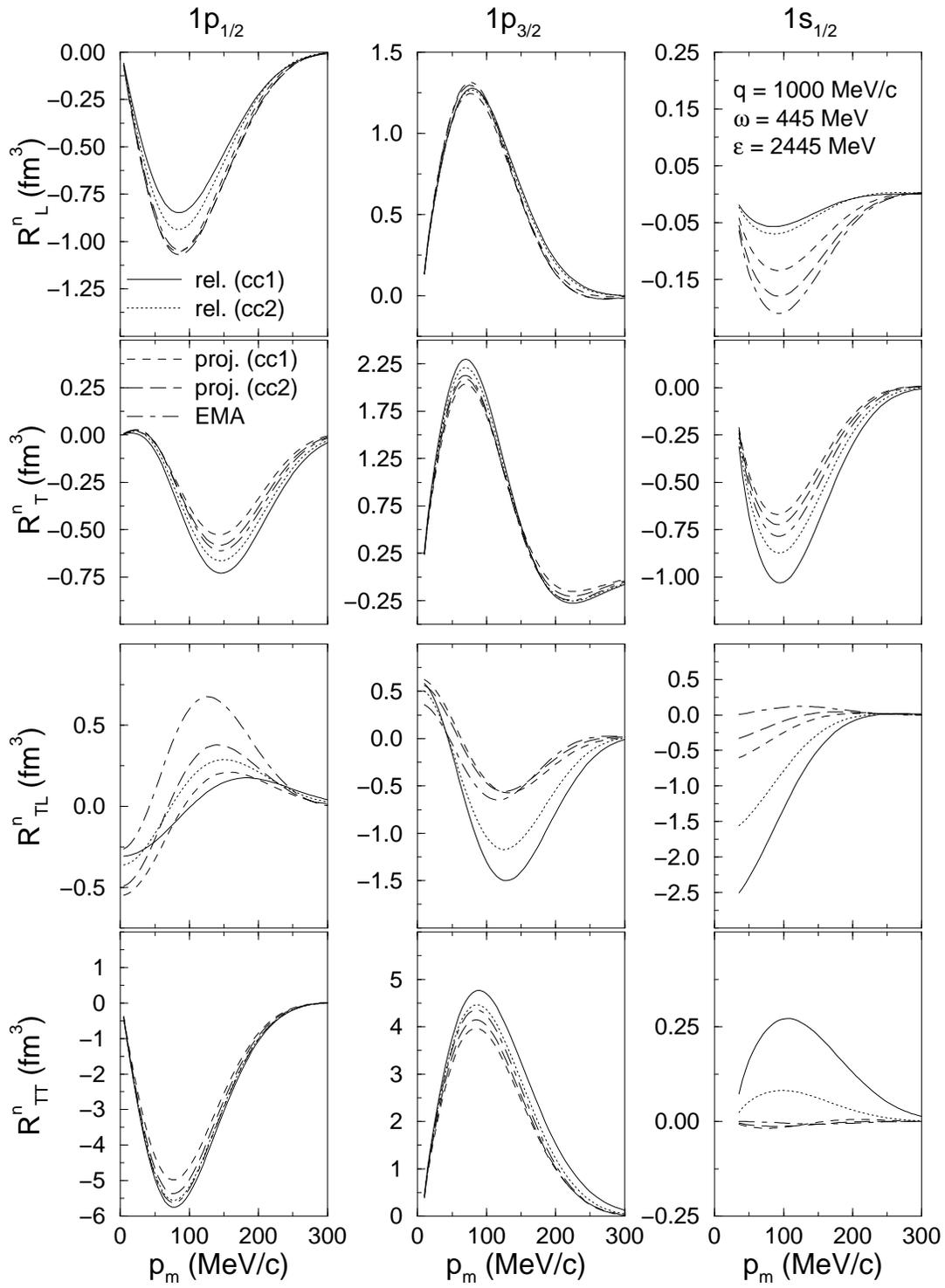,width=1.\textwidth}}
\end{center}
\caption{Normal responses for the kinematics of Figure 8.}
\end{figure}

\begin{figure}[t]
\begin{center}
\leavevmode
\mbox{\epsfig{file=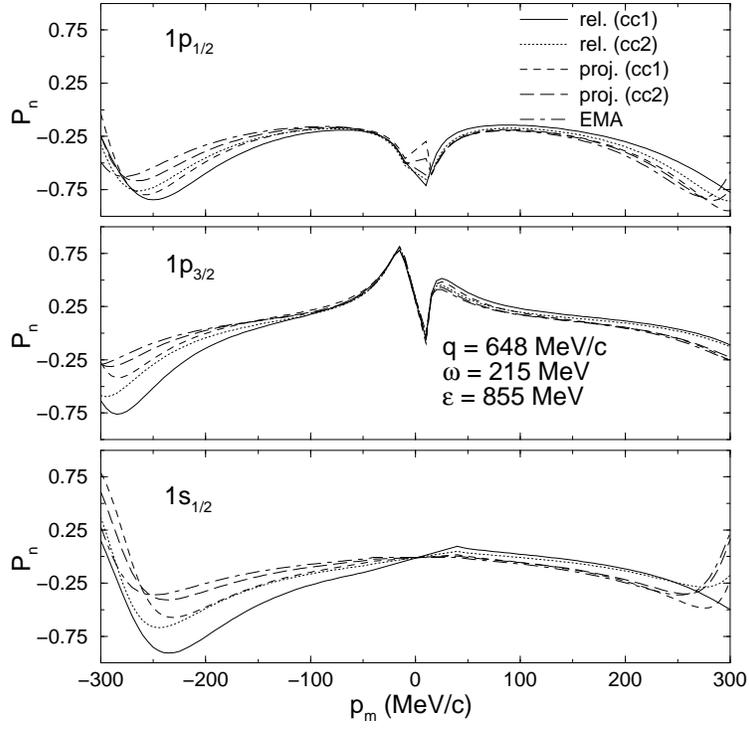,width=0.6\textwidth}}
\end{center}
\caption{The same as Figure 8 except for slightly smaller values of $q$ and
$P'$ as expected at Mainz \protect\cite{MAMI}.}
\end{figure}
\end{document}